\documentclass[11pt,a4paper]{article}

\usepackage[dvipsnames]{xcolor}
\pagecolor{white}

\usepackage{geometry}
\usepackage[utf8]{inputenc}
\usepackage{comment}%\begin{comment} ... comment lines ...  \end{comment}
\usepackage{amsmath}
\usepackage{amsfonts}
\usepackage{amssymb}
\usepackage[final]{graphicx}
\usepackage{graphicx}
\usepackage{subcaption}
\usepackage{cite}
\usepackage{hyperref} %collegamento ipertestuale nelle referenze
\usepackage{cleveref}
\hypersetup{colorlinks,bookmarksopen,bookmarksnumbered,
citecolor=red,
linkcolor=red,
pdfstartview=false,
urlcolor=blue}

\newcommand{\ket}[1]{|#1\rangle}
\newcommand{\bra}[1]{\langle #1 |}

\renewcommand{\th}{\vartheta}

\newcommand{\expec}[1]{\langle #1 \rangle}
\title{XXZ ladder quenches}
\author{gianluca.lagnese }
\date{November 2020}

\begin{document}

\title{Quenches and confinement in a Heisenberg-Ising spin ladder}
\author{Gianluca Lagnese$^{1,2}$, Federica Maria Surace$^{1,3}$, M\'arton Kormos$^4$, Pasquale Calabrese$^{1,2,3}$\\
$^1${\small SISSA, via Bonomea 265, 34136 Trieste, Italy}\\
$^2${\small INFN, Sezione di Trieste, 34136 Trieste, Italy}\\
$^3${\small International Centre for Theoretical Physics (ICTP), Strada Costiera 11, 34151 Trieste, Italy}\\
$^4${\small MTA-BME Quantum Dynamics and Correlations Research Group,}\\
{\small Budapest University of Technology and Economics, Budafoki \'ut 8., 1111 Budapest, Hungary}
}

\date{}

%		\maketitle
	
\maketitle

\begin{abstract}
We consider the quantum quench dynamics of a Heisenberg-Ising spin ladder which is an archetypal model in which confinement of elementary excitations is
triggered by internal interactions rather than an external field. 
We show that the confinement strongly affects the light cone structure of correlation functions providing signatures of the velocities of the 
mesons of the model. 
We also show that the meson masses can be measured from the real time analysis of the evolution of the order parameter.     
\end{abstract}

\section{Introduction}
\label{sec:intro}

The idea of confinement originally emerged in the context of quantum chromodynamics (QCD). 
Due to strong interaction, quarks are never observed isolated in nature, as they form bound states known as mesons and baryons \cite{Wilson1974}.  
The intrinsic non-perturbative nature of confinement as well as the difficulty to perform numerical simulations of strong interactions on classical devices 
challenge our ability to probe confinement in QCD. 
Actually,  insights into this fascinating phenomenon can be gained realising that some condensed-matter systems display a similar phenomenology. 
McCoy and Wu \cite{McCoy1978} were the first to understand that the confinement is also present in the Ising model, a paradigmatic model of statistical physics, 
in the ferromagnetic phase and in the presence of an external magnetic field.  
%This observation started an increasingly large number of theoretical investigations. 
Focusing on the transverse field Ising spin chain, in the ferromagnetic phase, the fundamental excitations are domain walls interpolating between the two ferromagnetic ground states. 
When a (even very small) longitudinal field is introduced, an effective attractive potential is generated and the domain walls get confined in bound states whose nature
is very similar to the mesons of QCD. 
Following the pioneering work by McCoy and Wu, confinement has then been found and investigated  with various analytical and numerical methods 
in thermal equilibrium in several quantum one-dimensional models  \cite{Delfino1996,DELFINO1998675,Rutkevich1999,Fonseca,Bhaseen2004,Rutkevich2005,Rutkevich2008,Rutkevich2010,r-17,gian20,rlxp-20,vwc-20}.

%CiracGaussian,,Brenes2018,PaiPretkoFractonsLGT missing!!

Fundamental results about confinement in condensed-matter systems have been achieved through the experimental study of various compounds (i.e. KCuF${}_3$\cite{tennant1993,tennant1995,satija1980}, Sr${}_2$CuO${}_3$\cite{ami1995} or YbAs${}_3$ \cite{schmidt1996}) that can be modelled as one-dimensional systems with a certain amount of inter-chain interaction. These studies on spin ladders were partially motivated by Haldane's prediction of a gap in the excitation spectrum for integer-$S$ antiferromagnets \cite{Hald1983}, as well as by the work of Shiba \cite{Shiba1980} who suggested that a weak inter-chain coupling between two spin 1/2 Heisenberg chain can explain the occurrence of discrete lines in the Raman spectrum of CsCoCl${}_3$ and CsCoBr${}_3$ \cite{yelon1975,achiwal1969}. Consequently, quasi one-dimensional spin ladders experienced an increasing research activity and the occurrence of confinement in these models has been intensively studied \cite{Hida1991,Shelton1996,Schulz1996,Essler1997,Sandvik, Nersesyan1997,affleck1998soliton,Augier1999,Greiter2002,Greiter2002a,Jung2006,Steinigeweg2015,Bera2017,Tonegawa2018,Rutkevich2018,Suzuki2018,Fan2020}, finally leading to remarkable observations through neutron scattering and high-resolution terahertz spectroscopy
\cite{Lake2010,Morris2014,Grenier2015,Wang2015b,Wang2016,Wang2018,Wang2019}.

In recent years, it has been realised that confinement presents distinct signatures even in non-equilibrium situations in real time \cite{cai2012,marton2017}.
This observation stimulated a cascade of theoretical activity on the subject \cite{jkr-19,rjk-19,Liu2018,Verdel19_ResonantSB,ch2019confinement,mazza2019,lerose2019quasilocalized,Castro-Alvaredo2020,mrw-17,cr-19,vcc-20,sl-21,kzhh-21,pwzt-21,bbm-21,lsmc-21,pp-20}, 
leading to a remarkable trapped ion experiment \cite{tan2019observation}, where confinement was probed in a non-equilibrium  
transverse field Ising model (TFIM) with long range interactions. 
Also, it was very recently observed on a IBM quantum computer \cite{vovrosh2021} simulating a TFIM with external longitudinal field (the same model employed in \cite{marton2017}). Furthermore, this research line brought fresh ideas to lattice gauge theories \cite{mdfppe,Hebenstreit2013,Pichler:2016it,Martinez2016,Surace}.
 
 The archetypal protocol to drive a system out of equilibrium is the quantum quench \cite{calcard06,calcard07}, which is the focus of this manuscript, too. 
The protocol is usually configured as follows: at time $t=0$ an isolated system is prepared in a pure state $\ket{\psi_0}$. Often, it is the ground state of a many-body Hamiltonian $H_0$. A Hamiltonian parameter is then suddenly changed and the system is evolved for $t>0$ as $\ket{\psi}=e^{- i H t} \ket{\psi_0}$ with a Hamiltonian $H$ that does not commute with $H_0$.  
The time evolution after a quantum quench is typically characterised by lightcone spreading of correlators \cite{calcard06,calcard07} and by a linear growth of 
entanglement in time \cite{calcard05}. 
Remarkably, confinement of elementary excitations fundamentally alters this picture: the light cone spreading and the growth of entanglement are 
highly suppressed~\cite{marton2017}. 
Furthermore, while typically the order parameter decays exponentially \cite{calcard06}, in a confining scenario it oscillates with frequencies related to the masses of the 
mesons~\cite{marton2017,vcc-20}. 
Recent works showed that confinement may lead to the absence of thermalisation \cite{ch2019confinement,cr-19}, 
to the emergence of rare non-thermal states ({\it scars}) in the many-body spectrum \cite{jkr-19,rjk-19},
and to fracton dynamics \cite{pp-20} as well as to anomalously slow dynamics \cite{mazza2019,lerose2019quasilocalized}. 
All these works addressed the dynamics of spin chains. 
Here, instead, we are interested in studying the dynamics in a confining spin ladder. 
In a ladder geometry the composite excitations (mesons) have a richer variety than in purely one-dimensional systems. 
As we shall see, this allows for more exotic non-equilibrium phenomena characterised by different energy scales. 

The paper is organised  as follows. In Sec.~\ref{sec:model} we describe the model, its relevant features, and the structure of its elementary excitations. Afterwards, in Sec~\ref{sec:quench} we introduce the quench protocol and we study numerically its non-equilibrium dynamics. 
More precisely, the effects of confinement in the model are displayed through the behaviour of the entanglement entropy evolution in Subsec.~\ref{sec:entanglement}, 
through the spreading of  two-point correlation functions in Subsec.~\ref{sec:lightcones}, and in the oscillations of the order parameter Subsec.~\ref{sec:stagmag}. We give our conclusions in Sec. \ref{sec:concl}.

 \section{The model}
\label{sec:model}

Here we study the non-equilibrium dynamics of a quasi one-dimensional model (ladder)
in its anti-ferromagnetic gapped phase. The model is realised as two
Heisenberg XXZ spin-($1/2$) chains coupled via an Ising interaction along the longitudinal direction \cite{Bhaseen2004,gian20}
\begin{equation}
\label{eq:ladder}
	    H(\Delta_{||},\Delta_{\perp}) = \frac{J}{2} \sum_{j=1}^{L}
	    \sum_{\alpha=1,2}\left[ \sigma^x_{j,\alpha} \sigma^{x}_{j+1,\alpha} +
	    \sigma^{y}_{j,\alpha} \sigma^{y}_{j+1,\alpha} + \Delta_{||}(\sigma_{j,\alpha}^{z} \sigma_{j+1,\alpha}^{z} + 1)\right]+ J\Delta_{\perp}\sum_{j=1}^{L} \sigma^{z}_{j,2} \sigma^{z}_{j,1}\,,
\end{equation}
where the operators $\sigma_{j,\alpha}^{x,y,z}$ are the Pauli matrices for the $j$-th spin of the $\alpha$-th chain. From now on we set the coupling to $J=1$. The anisotropy parameter $\Delta_{||}$ is chosen in the interval $\Delta_{||}\in (1, +\infty)$ such that the two chains are in their antiferromagnetic phases. 
We will take, without any loss of generality, $\Delta_{\perp}>0$, which favours anti-alignment of spins along the rungs of the ladder.

In the limiting case of zero inter-chain coupling ($\Delta_{\perp}=0$) the physics of the model is reduced to that of two independent XXZ anti-ferromagnetic spin chains. 
Each single XXZ chain has, in the strong anti-ferromagnetic limit $\Delta_{||} \to +\infty$, two degenerate ground states: N\'eel  $|\Psi_{1}\rangle = |\uparrow\downarrow\uparrow\downarrow\dots\rangle$ and anti-N\'eel $|\Psi_{2}\rangle = |\downarrow\uparrow\downarrow\uparrow\dots\rangle$ 
(here $\ket{\uparrow}$ is chosen with quantisation  axis in the z direction, i.e., $\sigma_j^z \ket{\uparrow} = \ket{\uparrow}$).
The fundamental excitations are domain walls toggling between  $|\Psi_{1}\rangle$ and $|\Psi_{2}\rangle$. Depending on the orientation of the neighbouring  spins, each domain wall can carry spin $s=+1/2$ (in the case $\uparrow \uparrow$) or $s=-1/2$ (if $\downarrow \downarrow$). Their dispersion relation, in the $\varepsilon=1/\Delta_{||} \to 0$ expansion, reads
\begin{equation}
\label{eq:omega}
    \omega(p)=1-2 \varepsilon \cos{2 p}.
\end{equation}

Away from the strong anti-ferromagnetic limit, for finite values of $\Delta_{||}$, the XXZ spin chain is integrable and can be solved via Bethe ansatz \cite{korepinbook}. 
When $\Delta_{||}\in (1, +\infty)$ the ground states are two N\'eel ordered states $|\Psi_{+}\rangle$ and $|\Psi_{-}\rangle$ with finite staggered magnetisation  $\pm \Bar{\sigma}$ given by
\begin{equation}
\label{eq:stagmag}
\bar{\sigma} = \prod_{n=1}^{\infty} \left( \frac{1-e^{-2n\gamma}}{1+e^{-2n\gamma}} \right) ^{2}\,, 
\qquad   \Delta_{||}=\cosh(\gamma )\,.
\end{equation}
The fundamental excitations in this case are charged quasi-particles called kinks interpolating between those two ground states. Like their $\Delta_{||}\rightarrow +\infty$ counterpart, they carry half-integer spin $s=\pm 1/2$. However, due to the properties of the model, they are strongly interacting particles with non-trivial dispersion relation and scattering phase.  
Their dispersion relation reads~\cite{zabrodin1992}
\begin{equation}
\label{eq:betheomega}
    \omega(\th)= \frac{2 K(k)}{\pi} \sinh{\gamma \sqrt{1-k^2 \cos^2 \th}}\,,
\end{equation}
where $K(k)$ is the complete elliptic integral of the first kind, the modulus $k$ is related to the anisotropy by the relation $K(\sqrt{1-k^2})/K(k)=\gamma /\pi.$

\begin{figure}
    \centering
    \includegraphics[width=0.8\linewidth]{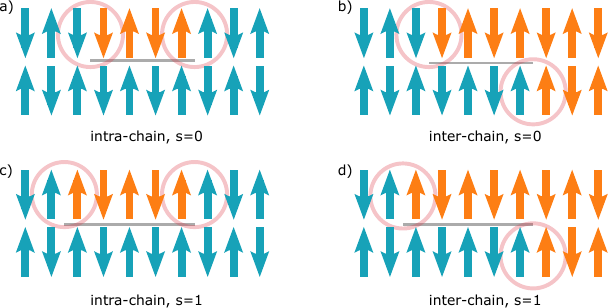}
    \caption{Illustration of inter-chain and intra-chain mesons. The colours blue/orange represent the two ground states $\ket{\Psi_{+-}}, \ket{\Psi_{-+}}$.}
    \label{fig:illustration}
\end{figure}

The spectrum of excitations in the spin ladder \eqref{eq:ladder} has been discussed in Ref. \cite{gian20} which we briefly summarise here.   
Because each chain is in an antiferromagnetic phase with two degenerate ground states, the ladder made of two decoupled spin chains has a four-fold degenerate ground state manifold 
($\ket{\Psi_{++}}$, $\ket{\Psi_{+-}},\ket{\Psi_{-+}}$, $\ket{\Psi_{--}}$, where the shorthand stands for $\ket{\Psi_{ij}}= \ket{\Psi_{i}}_1 \otimes \ket{\Psi_{j}}_2$ with $i,j=\pm$ and 
$1,2$ referring to the two chains) 
that is split by an inter-chain coupling $\Delta_{\perp}\neq 0$. 
For $\Delta_{\perp}\gtrsim 0$, $\ket{\Psi_{+-}}$ and $\ket{\Psi_{-+}}$ are the two degenerate ground states, whereas $\ket{\Psi_{++}}$ and $\ket{\Psi_{--}}$ acquire an extensive 
energy gap $\sim L \Delta_{\perp}$ on top of the ground state (for $\Delta_{\perp}< 0$ the opposite is true). 
In other words, the original $\mathbb{Z}_2\times \mathbb{Z}_2$ symmetry is explicitly broken by the inter-chain coupling to a single $\mathbb{Z}_2$ symmetry: 
this residual symmetry is spontaneously broken, leading to a two-fold degeneracy in the ground state. In this model, in analogy with the well-known mechanism leading to confinement in the Ising model in transverse and longitudinal field, the explicitly broken symmetry is responsible for the confinement of the excitations (kinks) into bound states called mesons. The mesons can be made of kinks located on the same or on different chains (see Fig. \ref{fig:illustration} for a pictorial illustration). While in the first case (intra-chain mesons) they toggle between the same type of ground state, in the second case (inter-chain mesons) they toggle between different ground states: similarly to kinks in a single chain, inter-chain mesons are topological excitations and can be created by local operations only in pairs.  Both types of mesons can either have spin $s=0$ or $s=\pm 1$.
%depending on the topological charge of the two kinks that are involved.

The illustration in Figure \ref{fig:illustration} %scenario showed in the picture is 
%the proper one
is a faithful picture of the mesons only in the limit $\Delta_{||} \to \infty$. Nonetheless, it holds qualitatively 
as long as we are in the gapped anti-ferromagnetic phase $\Delta_{||}>1$. Thus, the low-energy sector of the spectrum of the model is dominated by those two kinds of bound states. Their properties were investigated in Ref.~\cite{gian20} through numerical exact diagonalisation and approximate techniques, including a Bohr--Sommerfeld semiclassical approximation (in the following dubbed ``BS'') and the diagonalisation of the effective two-kink Hamiltonian (in the following dubbed ``Bessel''). We recall that the latter approach is accurate in the limit $\Delta_{||} \to \infty$. Conversely, the BS method is able to take into account the dressing of the elementary excitations when $\Delta_{||}$ is finite through their Bethe Ansatz dispersion relation and scattering matrix.  
The Bessel method consists in projecting the Hamiltonian~\eqref{eq:ladder} in an effective two-body Hamiltonian describing the motion of two kinks over the 
N\'eel ordered ground state. The solutions for the wave function are written in term of Bessel functions (and hence the name). 
Once diagonalised, the effective description gives the spectrum of the two-kink bound state for a given center of mass momentum $P$, which represents 
the meson dispersion relation. 
Likewise, although it works only for intra-chain mesons, the BS approach gives a center of mass momentum dispersion relation of the meson. 
It consists of applying the Born-Sommerfeld quantisation prescription to a system of two kinks propagating with dispersion relation Eq.~\eqref{eq:betheomega} and constrained 
by a potential $\propto \Delta_{\perp}$ growing linearly with the relative distance. 
The non-trivial interaction properties, neglected in the other approach, are encoded in the scattering phase shift.
For more details on these calculations, the interested reader can consult Ref.~\cite{gian20}. 

%%%%%%%%%%%%%%%%%%%%%%%

\section{Quench dynamics}
\label{sec:quench}

In this section we report the results of numerical simulations for the time evolution after a quench obtained by using the infinite volume Time Evolving Block Decimation (iTEBD) 
algorithm~\cite{itebd}. 
The system is prepared in a product state of the two chains with one chain in the
N\'eel $|\Psi_{1}\rangle = |\uparrow\downarrow\uparrow\downarrow\dots\rangle$ and the other in the anti-N\'eel state $|\Psi_{2}\rangle = |\downarrow\uparrow\downarrow\uparrow\dots\rangle,$ i.e., in the ground state of the model in the limit $\Delta_{||} \to \infty$, $\Delta_{\perp} \geq 0$. Thereafter, the quench is realised by letting the system evolve under the Hamiltonian with finite $\Delta_{||}>1$ and $\Delta_\perp>0$ (i.e., by suddenly changing both the inter-chain and intra-chain couplings).
The real time evolution after the quench  is performed with a Trotter step $\delta t=10^{-2}$. The bond dimension $\chi$ is set to $512$. We checked the stability of the numerical simulations with respect to changes in $\chi$ and $\delta t$. 

In the following, we investigate the effects of confinement (i.e., of a non-zero the inter-chain coupling $\Delta_{\perp}$) on the evolution of the entanglement entropy, 
the one-point function of the order parameter, and its equal time two-point correlation function. 
For $\Delta_{\perp}\neq 0$ (where the model is not integrable) we will take advantage of the aforementioned results on the properties of the excitations 
reported in \cite{gian20} 
to interpret the dynamics.

\subsection{Entanglement entropy}
\label{sec:entanglement}

\begin{figure}[!t]
    \centering
    \includegraphics[width=\textwidth]{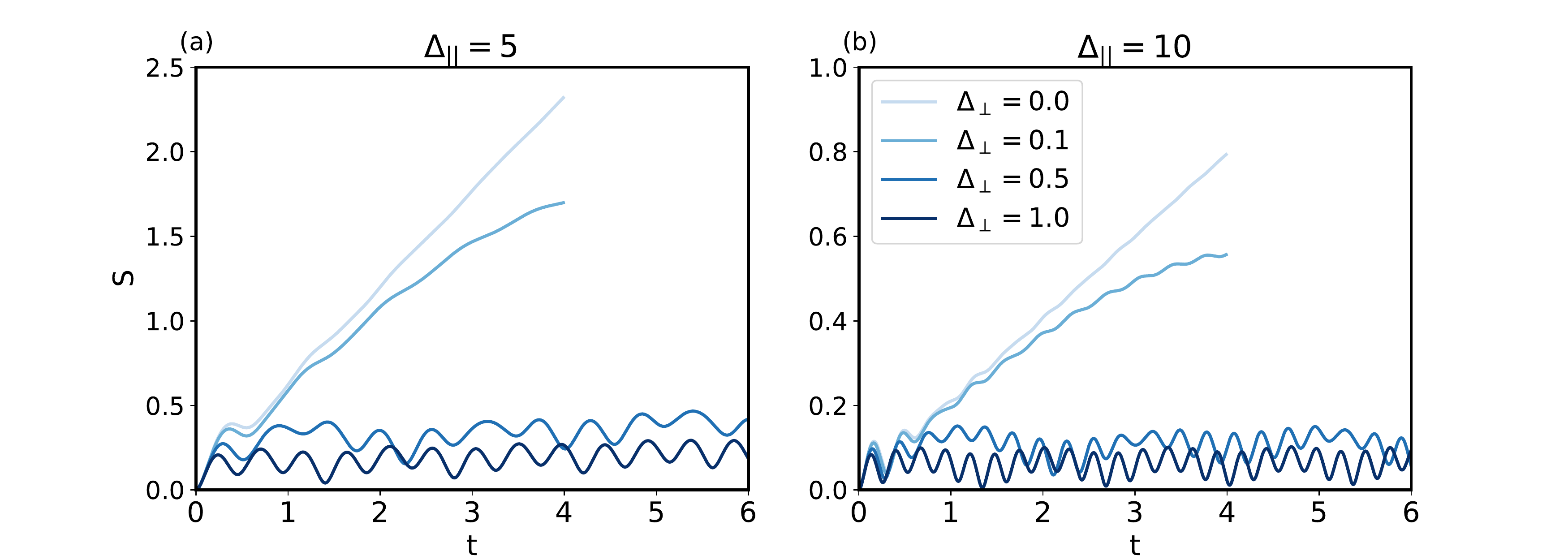}
    \caption{Time evolution of the half chain entanglement entropy \eqref{St} after a quench from a state with one chain in the N\'eel state and the other in the anti-N\'eel state, 
    i.e. the ground state for $\Delta_{||}=\infty$. The post-quench Hamiltonian has (a) $\Delta_{||}=5$  and (b) $\Delta_{||}=10$. 
    The different shades of blue correspond to different values of $\Delta_{\perp}$, as shown in the legend.  
    Note that the linear growth for $\Delta_\perp=0$ is turned into an oscillatory behaviour in the presence of confinement.}
    \label{fig:vnntropy}
\end{figure}

The entanglement entropy has been shown to encode a considerable amount of information about the out-of-equilibrium dynamics of many-body systems and is a probe for lightcone effects. Its time evolution after a quench evoked an intense research activity (see e.g. the review \cite{c-20} for a complete bibliography), 
and it is even accessible experimentally in cold atoms and ion traps \cite{kaufman16,brydges-2018,daley12,islam15,lukin-18,vek-21}. 

Here we focus on the half chain entanglement entropy, measured through the von Neumann entropy 
\begin{equation}
S(t)= -\mathrm{Tr} \left[ \rho_\text{h}(t) \log \rho_\text{h}(t) \right]  ,
\label{St}  
\end{equation}
of the half chain reduced density matrix $\rho_\text{h}$ obtained from the full density matrix $\rho(t)=\ket{\psi(t)} \bra{\psi(t)}$ by tracing out the degrees of freedom of the other half chain. The results are shown in Fig. \ref{fig:vnntropy}.

When the inter-chain coupling $\Delta_{\perp}$ is zero, the well-known linear growth of the half-chain
entanglement entropy is observed. This limiting case is essentially a standard quench in the two independent XXZ chains in the anti-ferromagnetic region, whose dynamics have been thoroughly studied in  Refs.~\cite{alba17,ac-18}. The dynamics of an integrable model, due to the infinite number of conserved quantities, may be understood in terms of quasi-particles complemented by the knowledge of the stationary states from Bethe Ansatz. 
Here the quasi-particle picture \cite{calcard05} provides an intuitive yet quantitative framework for quenches: 
the pre-quench state acts as a source of pairs of quasi-particles with opposite momenta. 
The pair of particles are entangled and, by traveling ballistically, they spread quantum correlations through the system. 
In an interacting integrable model there are different quasi-particle species, each one contributes to $S(t)$ proportionally to the number 
of quasi-particles shared between the two subsystems. 
Therefore, $S(t)$ is the sum over the independent contributions of each quasi-particle species. 
Since they travel ballistically, a pair of quasi-particles can spread correlations over a distance that grows linearly in time, leading to a linear growth $S(t)\propto t$. 
We observe this linear behaviour in Fig. \ref{fig:vnntropy} (top lines) for two different post-quench values of $\Delta_{||}$. 

A linear entanglement growth is generically expected also in non-integrable models, despite the lack of well defined quasi-particles
(see for instance  \cite{KimHuse2013,nahum17}).
In the model we are studying, the presence of the inter-chain Ising coupling $\Delta_{\perp}$ does break integrability. 
However, once the inter-chain coupling $\Delta_{\perp}$ is introduced, the growth of the entanglement entropy is significantly slowed down with the appearance of large oscillations that are stable within the observation time. 
The behaviour of the von Neumann entropy showed in Fig. \ref{fig:vnntropy} can be explained by the following argument. 
In contrast to kinks that can only appear in pairs, the excitations produced in the quench for $\Delta_{\perp}\neq 0$ mainly consist of single mesons: 
these mesons have zero momentum, and they contribute to entanglement through the mechanism of Bloch oscillations resulting in the periodic variation of the size 
of the meson  (``breathing''). 
Correlations can only spread roughly up to the distance set by the largest separation of the two kinks forming the meson. 
Therefore, the entanglement entropy grows linearly only for a limited time (independent of system size) and then bounce back oscillating.
The frequencies of the oscillations are expected to correspond to the meson masses and their differences (we do not report here the analysis 
of the frequencies of these oscillations because it is identical to the one below in Fig. \ref{fig:stagmag} for the one-point function).
%that mesons are mostly produced at rest.

Above the two-particle threshold, a small fraction of mesons are expected to be produced in pairs with non-zero and opposite momenta.
%at a nonzero velocity.
These pairs of mesons can propagate freely, leading eventually to a linear growth of entanglement entropy at long times.
However, the production of meson pairs is very small for the small quenches we studied and its effect on the entanglement is too slow to be detected numerically. However, as shown in Sec.~\ref{sec:lightcones}, some effects of the production of pairs of mesons are detectable in the time evolution of two-point correlations.

\begin{figure}[!t]
    \centering
    \includegraphics[width=\textwidth]{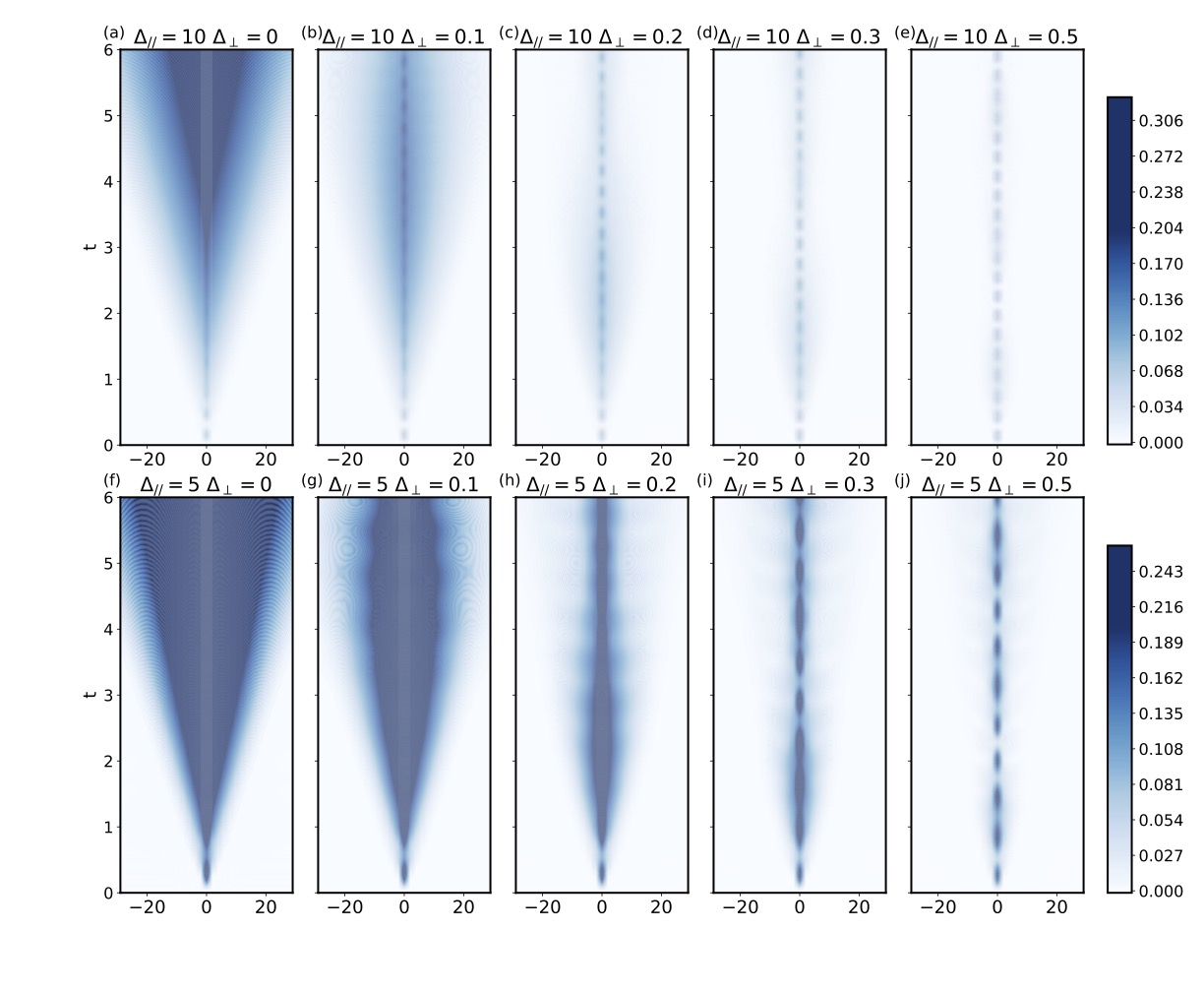}
 \caption{\label{fig:lightcones}  Space-time plots of the time evolution of the single-chain connected staggered two-point function $C_s(\ell)=\expec{(-1)^{\ell}S^z_{i,1}S^z_{i+\ell,1}}_c$ 
 after a quench. The initial state has one chain in a N\'eel state and the other in anti-N\'eel state. 
 The final Hamiltonian has  $\Delta_{||}=10$   (top row)
 %$\Delta_{\perp}=0,0.1,0.5$ in (a,b,c) 
 and $\Delta_{||}=5$ (bottom row)  with $\Delta_{\perp}=0,0.1,0.2,0.3,0.5$. 
 The standard ballistic light cones for $\Delta_\perp=0$ (a,f) are strongly suppressed by turning on a non zero $\Delta_\perp$.}
\end{figure}

\subsection{Light cones in the two-point function}
\label{sec:lightcones}

In this section we investigate the effect of the inter-chain coupling $\Delta_{\perp}$ on the spreading of two-point correlation functions. 
We focus on the  equal time connected two-point function of the local staggered magnetisation, i.e. 
$C_s(\ell) = \expec{(-1)^{\ell}S^z_{i,\alpha} S^z_{i+\ell, \alpha} }_c= \expec{(-1)^{\ell}S^z_{i,\alpha} S^z_{i+\ell, \alpha} }- \expec{(-1)^{i}S^z_{i,\alpha}}^2$.  
Because of the Lieb--Robinson bound \cite{liebrobinson72},  there is a maximal velocity of propagation $v_\text{max}$. % \geq v(\lambda)$.
Consequently, at time $t$ all the connected correlators vanish at distances $\ell \geq v_\text{max} t$. 

Let us start by considering $\Delta_{\perp}=0$ where, like for the other observables, the behaviour of two-point connected correlators can be understood in terms of the 
quasi-particle description of the XXZ chain: the quasi-particles travel ballistically with velocities $v(\lambda)<v_\text{max}$, leading to the light cones observed in   
Fig. \ref{fig:lightcones}-(a),(f). The velocity $v_{\rm max}$ is related to the maximum speed of excitations built on top of the stationary state, see Ref. \cite{bonnes-2014}.
As pointed out in Ref. \cite{marton2017} for the Ising chain, in the presence of confinement the quasiparticles (i.e. the kinks) cannot move anymore ballistically and they 
show a characteristic breathing shape with an amplitude of the oscillation related to width of the zero-momentum meson.
This transition from the ballistic growth to the confined regime is reported in Fig. \ref{fig:lightcones} showing clearly that when $\Delta_{\perp}\neq0$,  the quasiparticles 
are confined into mesons. 
As a consequence the correlations are strongly suppressed and they are considerably non-zero only in the region where the mesons extend, see again 
Fig. \ref{fig:lightcones}.
%\gl{light cone has been confirmed through exact calculations and numerical simulations as well as experimentally confirmed...?}

\begin{figure}[!t]
    \centering
    \includegraphics[width=\textwidth]{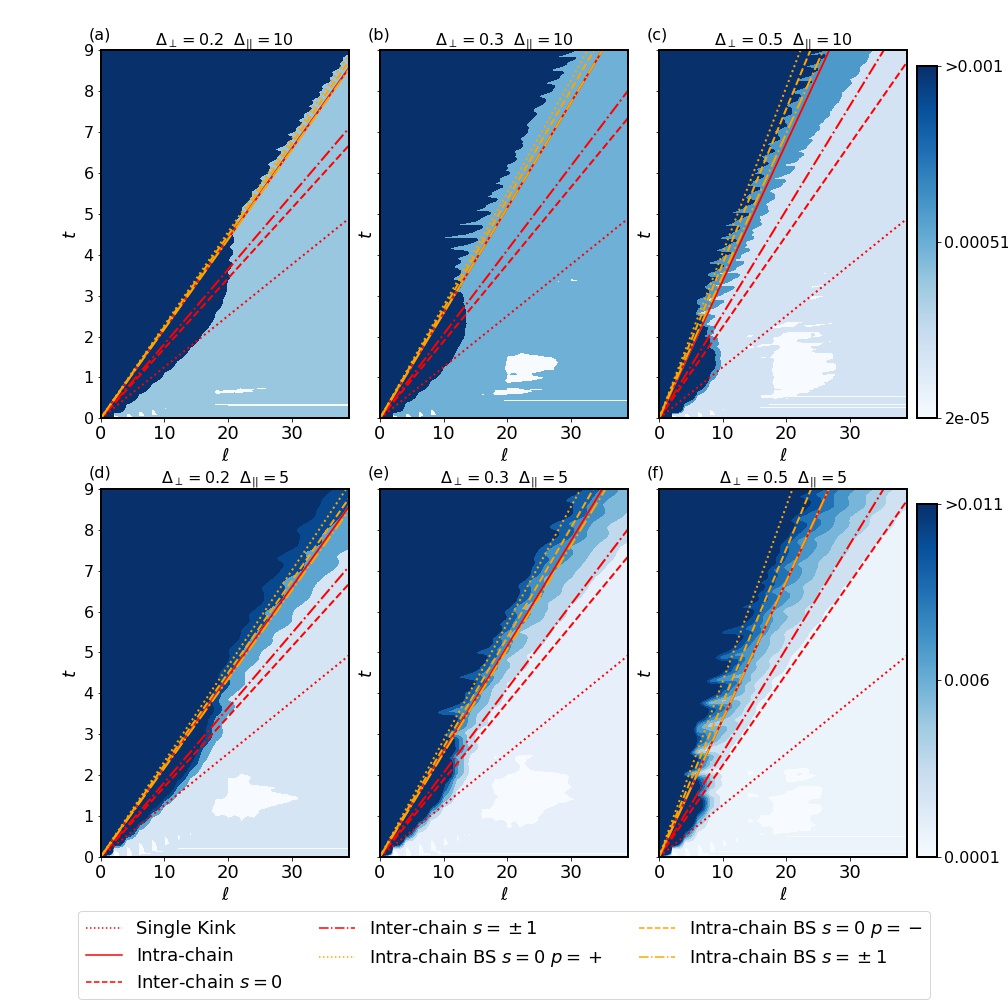}
 \caption{
Density plot for time evolution of the single-chain connected staggered two-point function $C_s(\ell) = \expec{(-1)^{\ell}S^z_{i,1} S^z_{i+\ell,1} }_c$ after a quench. 
The initial state has one chain in the N\'eel state and the other in the anti-N\'eel state. The post-quench Hamiltonian has $\Delta_{||}=10$ with $\Delta_{\perp}=0,0.1,0.5$ in (a), (b) and (c) respectively and $\Delta_{||}=5$ with $\Delta_{\perp}=0,0.1,0.5$ in (d) , (e) and (f). 
The scale in the colour plot is tuned in such a way that a weaker signal compared to Fig. \ref{fig:lightcones} is visible. 
The straight lines correspond to the maximal speed of propagation of the different particles. 
The red dotted line corresponds to the kink propagation speed, obtained from Eq.~\eqref{eq:betheomega}. 
The other lines describe the propagation of the two kinds of mesons. 
All yellow lines are BS predictions while the red ones are Bessel.
In the Bessel approximation, all intra-chain mesons have the same maximal speed (that is why we report a single curve).
These are resolved by the BS approach showing two lines, one for spin $s=0$ and negative parity the other for spin $s=\pm1$.
Inter-chain mesons are instead found only in the Bessel approximation.
%The yellow dashed are the BS prediction for the intra-chain meson. Bessel approach is used for all the others. 
%They correspond to the various mesons as reported in the bottom legend. 
 }
\label{fig:lightcones2} 
\end{figure}

%%%%%%%%%%%%%%%%%%%%%%%%%%%%%%%%%%%%%

Till now the form of the correlations does not show qualitative differences compared to the confining Ising chain reported in Ref. \cite{marton2017}.
In order to observe the effect of the presence of different species of mesons, in Fig. \ref{fig:lightcones2} 
we re-plot some of the density plots of Fig. \ref{fig:lightcones} on a different scale (with respect to Fig.~\ref{fig:lightcones}, being the figure symmetric, 
we consider the half plots with  $\ell>0$).
In this way we can appreciate the small signal due to the production of pairs of mesons of opposite momenta: weak components of the signal, previously hidden, become visible
showing several light cones.
The signal clearly shows that these secondary light cones are a consequence of  propagating mesons:
as these are heavier particles compared to the kinks, they move at a lower velocity. 
To understand quantitatively their structure, we compare them with the maximal speed of propagation of the different mesons and of the kinks.
These velocities are extracted with the help of the approximate techniques developed in Ref. \cite{gian20} 
(and briefly mentioned at the end of Sec. \ref{sec:model}) for the derivation of the dispersion relations $\omega_m$, where $m$ runs on the kind, spin and parity of the 
meson (for details see \cite{gian20}).
The maximal speed of propagation of the a given meson is then obtained as $v^{(m)}_\text{max}=\max\left(\frac{\mathrm{d}\omega_m(p)}{\mathrm{d}p},p \in [-\pi/2,\pi/2]\right)$  
(the derivative of the dispersion relations is computed numerically with a proper discretisation of the center of mass momentum $P$). 
In the figure we report all the speeds of propagation of the lightest (i.e. fastest) meson for each type (details in the caption).
We compare them to the kink speed, expected to describe the $\Delta_{\perp}=0$ case (and obtained from Eq.~\eqref{eq:betheomega}).

Let us now finally discuss the results are shown in Fig.~\ref{fig:lightcones2} for two values of the intra-chain coupling $\Delta_{||}=5,10$ 
and inter-chain coupling $\Delta_{\perp}=0.2,0.3,0.5$.
%The lightcone propagating speeds are the ones associated to the lightest (the fastest) mesons. 
Right after the quench, the evolution of the light cone is always compatible with the one predicted from the free kink dispersion relation.  
Once the two kinks are created with opposite momenta, they do not immediately feel the confining potential and they experience an almost-free initial propagation. 
Subsequently, they bounce back, and the signal deviates significantly from the free slope. 
The main signal is associated to zero momentum intra-chain mesons which corresponds to the breathing observed in Fig. \ref{fig:lightcones}.
Outside of the first meson breathing zone, secondary light cones develop which are visible on the scale of Fig. \ref{fig:lightcones2}. 
In all the cases that are shown, the first secondary cone (the darkest one in the figure) has a slope which is compatible with 
the maximal speed of propagation associated intra-chain mesons.
For this speed, we report the derivation with the two approaches of Ref. \cite{gian20} (yellow and red curves) although the
differences are very small and the numerical data (accessible in the observation time) are unable to discriminate among the two. 
The other secondary light cones (the one in lightest blue) are expected to be explained in terms of lighter mesons. 
In particular, some signal from pairs of propagating inter-chain mesons are also expected to be present. 
Though our results are not conclusive, in (c),(d),(e) and (f) a weaker signal is deviating from the main secondary lightcone and it is roughly compatible with the 
slopes we predicted for the inter-chain mesons.

\begin{figure}[t]
    \centering
    \includegraphics[width=\textwidth]{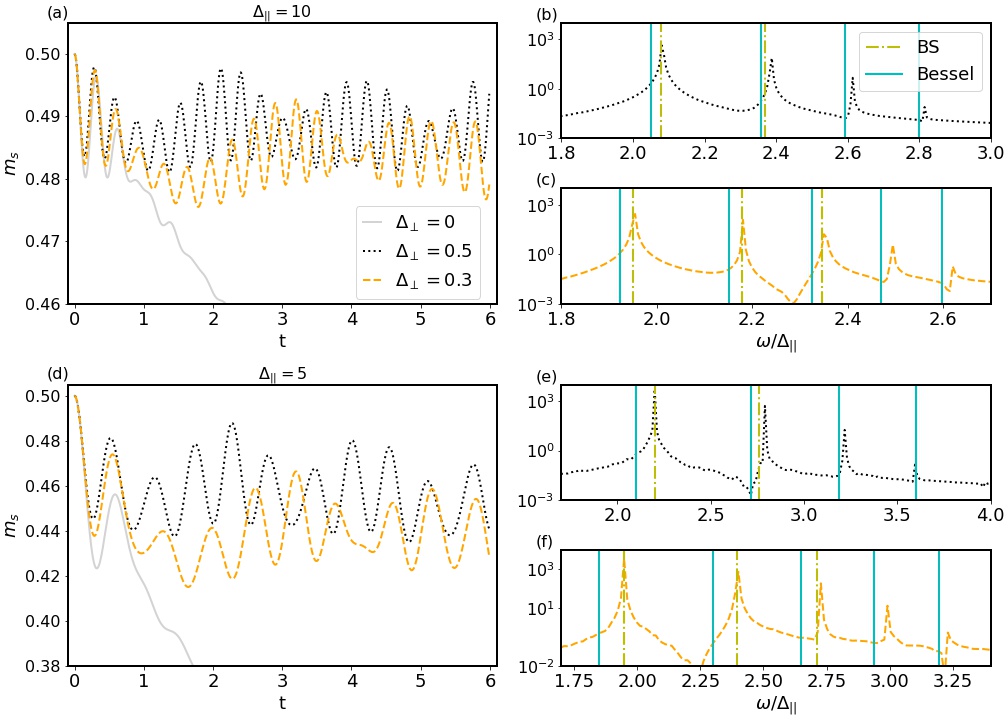}
    \caption{\label{fig:stagmag} Time evolution of the staggered magnetisation after a quench from a state with one chain in a N\'eel state and the other in Anti-N\'eel state.
    The final Hamiltonian has $\Delta_{||}=10$ (a) and $\Delta_{||}=5$ (d) with $\Delta_{\perp}=0.5$ (dotted) or $\Delta_{\perp}=0.1$ (dashed). 
    The light grey line is the quench to $\Delta_{\perp}=0$. 
    (b) and (c) represent the Fourier analysis relative to $\Delta_{||}=10$ for $\Delta_{\perp}=0.5$ and $\Delta_{\perp}=0.1$ respectively. 
    Likewise, (e) and (f) show the Fourier analysis relative to $\Delta_{||}=5$. 
    In (b) (c) (e) and (f) the peaks are compared with the prediction extracted from the semi-classical approximation (BS) (dot-dashed lines) and the exact diagonalisation 
    of the effective two-body Hamiltonian (Bessel) (straight lines). }
\end{figure}

\subsection{Time evolution of the staggered magnetisation and masses of the mesons}
\label{sec:stagmag}

After the entanglement entropy and correlation functions, we turn to the study of 1-point functions of the staggered magnetisation of a single chain $m_s=\expec{(-1)^i S_{i,1}^z}$, 
which is a local order parameter of the model in the anti-ferromagnetic region. 
The numerical results are presented in Fig. \ref{fig:stagmag} for the post-quench couplings $\Delta_{||}=10$ in (a),(b) and (c); 
$\Delta_{||}=5$ in (d) (e) and (f). %at $\Delta_{\perp}=0,0.1,0.5$ as explained in the legend.
For $\Delta_{\perp}=0$ (i.e. quenching $\Delta_{||}$ in an XXZ chain), the anti-ferromagnetic order is expected to relax exponentially \cite{calcard06,calcard07,bpg-10}, as showed by the solid grey line in the (a) and (d) panels. For $\Delta_{\perp} \neq 0$, instead, the staggered magnetisation  is trapped in stable oscillations, 
as for the Ising chain \cite{marton2017} (see also \cite{delfino-14,dv-14}). 
In the right hand side of Fig.~\ref{fig:stagmag}, (i.e. (b),(c),(e) and (f)), we show the results in the frequency domain after performing a Fourier analysis of the signals obtained for $\Delta_{\perp}=0.3,0.5$. The dominant oscillation frequencies appear as well-defined peaks.  

In order to show that the peaks correspond to the mass gaps of the mesons, we display the results of two different approximations for the masses discussed in our previous 
work Ref.~ \cite{gian20} (and briefly mentioned in section \ref{sec:model}). We compare the numerical predictions with both BS and Bessel approaches. 
%With the reader's intuition supported by Fig.~\ref{fig:illustration}, 
To analyse the data we recall that the Hamiltonian in Eq.~\eqref{eq:ladder} preserves the parity of the state and the total spin along each chain. 
Consequently, the post-quench state is constrained to share those properties with the initial state. 
In our case, the latter has one chain in N\'eel and the other in an anti-N\'eel state, so the relevant mesons must have total spin $s=0$ along each chain of the ladder and must have 
positive parity. 
%It is then clear that in the zero momentum sector we can expect to observe only $s=0$ intra-chain mesons. 
%
In the right hand side panels of Fig. \ref{fig:stagmag} we report theoretical predictions for the $s=0$ intra-chain mesons. 
It is evident that the Bessel approach roughly captures the position of the peaks, but quantitatively is a bit off. 
This is not surprising since the very same conclusion was also drawn for the equilibrium data in Ref. \cite{gian20}. 
However, for those states for which we found a solution of the BS quantisation condition, we have an an extremely precise description of the peaks. 
Furthermore, the improvement of accuracy of the BS approach compared to Bessel is more appreciable for $\Delta_{||}=5$ than for $\Delta_{||}=10$. 
This is explained as the more the longitudinal coupling $\Delta_{||}$ is decreased the more the corrections due to the non-trivial two-kink scattering, 
(and the precise form of their dispersion relation) become relevant.
We finally mention that a similar spectroscopy analysis can be done also for the entanglement entropy, but the obtained results are completely equivalent to those
for the staggered magnetisation and so are not reported here.

\section{Conclusions}
\label{sec:concl}

In this manuscript we systematically characterised the quench dynamics of the Heisenberg-Ising ladder with Hamiltonian \eqref{eq:ladder} in the region
of parameters presenting confinement, i.e. in the ordered antiferromagnetic phase of the two chains for $\Delta_{||}>1$. 
We show that the through spectroscopy of the order parameter evolution (and also of the entanglement entropy not reported here) the masses of the $s=0$ intra-chain mesons 
can be accessed, see Fig. \ref{fig:stagmag}.
A very remarkable finding is that while the two-point function are strongly suppressed because of confinement (see Fig. \ref{fig:lightcones}), there are feeble secondary light cones 
(see Fig. \ref{fig:lightcones2}) in which the existence of all types of mesons can be observed as a consequence of a very small production of pairs of mesons with opposite momenta. 
Incidentally, it is possible that this hybridisation with multiparticle states could eventually allow the system to thermalise, but on extremely long times scales. 
It would be very interesting to find even approximate methods to argue whether this is the case.

\section*{Acknowledgements}
We thank G\'abor Tak\'acs for useful discussions and for granting access to the cluster at BME where part of the numerical calculations have been carried out.
GL thanks BME and MK thanks SISSA for hospitality.
We thank ERC for partial support under grant number 758329, AGEnTh, (FS) and 771536, NEMO, (GL and PC).
%MK acknowledges the Hungarian Quantum Technology National Excellence Program, project no. 2017-1.2.1- NKP- 2017-00001, and by the Fund TKP2020 IES (Grant No. BME-IE-NAT). 
MK acknowledges support by a Bolyai J\'anos grant of the HAS, and by the National Research, Development and Innovation Office (NKFIH) through the OTKA Grant K 138606, the grant TKP2020 IES Grant No. BME-IE-NAT, and by the \'UNKP-20-5 new National Excellence Program of the Ministry for Innovation and Technology.

\end{document}